\documentstyle[floats,prl,aps]{revtex}
\begin{document}

\newcommand{\tl}{\tilde}
\newcommand{\bm}{\boldmath}
\newcommand{\cut}{\rm cut}
\newcommand{\MNRAS}{Mon. Not. Roy. Astron. Soc.}
\def\bi#1{\hbox{\boldmath{$#1$}}}

\draft
\title{ Lensing effect on polarization in microwave background: 
extracting convergence power spectrum}

\author{Jacek Guzik\cite{jacekmail}}
\address{Astronomical Observatory, Jagiellonian University, 
         Orla 171, 30-244 Krak\'ow, Poland}

\author{Uro\v s Seljak\cite{urosemail}}
\address{
Department of Physics, Jadwin Hall, Princeton University,
     Princeton, NJ 08544 
}

\author{Matias Zaldarriaga\cite{matiasemail}}
\address{Institute for Advanced Studies, School of Natural Sciences, Princeton, NJ 08540}
\date{October 1999}
\maketitle
\begin{abstract}
Matter inhomogeneities along the line of sight deflect the 
cosmic microwave background (CMB) photons originating 
at the last scattering surface at redshift $z \sim 1100$.
These distortions modify the pattern of CMB polarization.
We identify 
specific combinations of Stokes $Q$ and $U$ parameters
that correspond to spin 0,$\pm 2$ variables and 
can be used to reconstruct the projected matter density. 
We compute the expected signal to noise 
as a function of detector sensitivity and angular resolution. 
With Planck satellite the detection would be 
at a few $\sigma$ level. Several times better detector sensitivity 
would be needed to 
measure the projected dark matter power spectrum 
over a wider range of scales, which could 
provide an independent confirmation of the projected matter 
power spectrum as measured from other methods.
\end{abstract}
\pacs{PACS numbers: 98.80.Es,95.85.Bh,98.35.Ce,98.70.Vc  \hfill}
\section{Introduction}
Cosmic microwave background (CMB) is believed to 
originate from the epoch when protons and electrons recombined 
into neutral hydrogen making the universe transparent, 
which happened around redshift $ z \sim 1100$.
Photons travelling along the line of sight towards the observer 
were deflected by intervening 
dark matter distribution via the gravitational lensing effect. 
This process conserves two quantities: surface brightness and 
polarization. Conservation of surface brightness implies that 
in the absence of any fluctuations none can be generated by 
gravitational lensing. However, if there are fluctuations 
present then gravitational lensing can either smooth these
out on large scales \cite{uroslens} or generate new fluctuations 
on small scales \cite{metcalf}. 
The effect of these 
distortions on the CMB anisotropies has been thoroughly
investigated, both on the 
CMB power spectrum 
\cite{uroslens,others} as well as on the induced nongaussian 
signatures \cite{nongauss}. These studies have shown that the lensing effect 
is small, but detectable with future CMB experiments, and 
would provide important information on the distribution of 
dark matter up to $z \sim 1100$. This method possibly measures clustering 
amplitude at higher redshifts and larger scales than any other method
and would be specially valuable in breaking the degeneracies between 
cosmological parameters present when one only uses the CMB data.

Given the potential importance of measuring the dark matter power 
spectrum on large scales and/or high redshifts, it is worth 
investigating other methods than can provide similar information.
This is particularly important because any such measurement will 
only be statistical in nature and could
be susceptible to various systematic effects.
One such method is gravitational lensing on the CMB polarization.
Polarization is also conserved 
by gravitational lensing so that the latter only  
moves the polarization tensor from one point 
to another conserving its amplitude and direction in the mapping. 
Since it is widely 
expected that CMB polarization will be detected with the future
CMB experiments it is worth asking what kind of information 
can be learned from investigating the lensing effect on polarization 
and how can it be extracted. 

The effect of lensing on the power spectrum of CMB polarization 
has already been explored in \cite{zs98}. It has been shown 
there that lensing smoothes the acoustic peaks of $E$ 
polarization and also generates a $B$-type polarization.
While this induced $B$ polarization is small, it provides 
a fundamental limit to the level of $B$ polarization from 
other sources that can be detected on large scales. 
For example, it limits the detectable tensor to scalar ratios 
to be above $10^{-4}$, although this is not a significant 
limitation for the near future.

In this paper we explore in more 
detail the prospects of directly detecting gravitational 
lensing effect on the CMB polarization through its nongaussian 
signatures. We show that 
there is a combination of derivatives of polarization that 
provide a local estimator of shear and convergence. These, 
while noisy, can be used to provide an estimator of projected 
dark matter power spectrum, which by averaging over all the sky 
may yield a detectable signal. We compute the expected signal 
to noise as a function of detector sensitivity and apply it 
to the future satellite experiments. 

\section{Convergence from the CMB polarization}
\label{sec1}

In this section we develop a method to extract the gravitational lensing 
effect on polarization in the ideal case without detector noise and 
beam smoothing. Next section will deal with these complications.
A radiation field can be fully
described in terms of four Stokes parameters \cite{transfer} or 
the specific intensity tensor $I_{ij}$.  
For the CMB the Stokes parameters can be chosen as the temperature anisotropy
$T$ proportional to the sum of intensities along two perpendicular directions 
in the sky\footnote{
We work in the small scale limit which simplifies the 
expressions. Since the final results indicate that lensing on 
polarization is not detectable on very large scales this does 
not impose a significant limitation on the present work.},
the Stokes parameter
$Q$, defined as the difference
between the intensities along the two axis 
and the Stokes parameter $U$, defined by the difference between intensities
along the two diagonals.
The fourth Stokes parameter $V$ 
describes circular polarization which is not generated via the Thomson
scattering believed to be the only generating mechanism for polarization,
so we will ignore it in the following.
 
The Stokes parameters $Q$ and $U$ are two fields in the sky, 
which we assume to be Gaussian random fields that are generated 
from a common scalar potential (we ignore any initial 
$B$-type plarization, which, even if present, would likely be important 
only on large scales).
Because gravitational lensing conserves polarization
their observed values in the ${\bi \theta}$ direction
$Q({\bi \theta})$ and $U({\bi \theta})$ are related to their
values at the recombination $\tl Q$, $\tl U$, 
deflected by an angle $\delta{\bi\theta}$
\begin{eqnarray}
  Q({\bi \theta}) & = & \tl Q(\bi{\theta}+\delta{\bi \theta}) \nonumber \\
  U({\bi \theta}) & = & \tl U(\bi{\theta}+\delta{\bi \theta}). 
\label{qu}
\end{eqnarray}
To extract the lensing information contained 
in these fields we consider their spatial derivatives. These can be written
in the weak lensing regime as 
\begin{eqnarray}
  Q_a({\bi \theta}) = (\delta_{ab}+\Phi_{ab})
                    \tl Q_b(\bi{\theta}+ \delta{\bi\theta})  \nonumber \\
  U_a({\bi \theta}) = (\delta_{ab}+\Phi_{ab})
                    \tl U_b(\bi{\theta}+ \delta{\bi\theta}), \label{dert}
\end{eqnarray}
where $\Phi_{ab}={\partial \delta \theta_a \over \partial \theta_b}$ 
with $a,b=x,y$ are the shear tensor components describing the lensing effect.

Just like in the temperature case \cite{letlens} we form quadratic combinations of 
derivatives of the Stokes parameters and express them
in terms of unlensed variables and the components of the shear tensor. 
We wish to 
begin with a quantity that does not depend on the coordinate frame, which 
allows us to simply relate 
the unlensed and the lensed quantities. Such a 
quadratic quantity is readily available and is given by 
$Q^2+U^2$. Then, in the lowest order, we have 
\begin{eqnarray}
{\cal S}_P & \equiv & 
\left[Q_x^2+Q_y^2+U_x^2+U_y^2\right]({\bi \theta}) \nonumber \\
	&=&(1+\Phi_{xx}+\Phi_{yy})\tl {\cal S}_P+(\Phi_{xx}-\Phi_{yy})
	\tl {\cal Q}_P +
	2 \Phi_{xy} \tl{\cal U}_P \nonumber \\
{\cal Q}_P &\equiv& \left[Q_x^2-Q_y^2 + U_x^2-U_y^2\right]({\bi \theta}) \nonumber\\
	&=&(1+\Phi_{xx}+\Phi_{yy})\tl {\cal Q}_P+(\Phi_{xx}-\Phi_{yy})
	\tl {\cal S}_P \nonumber \\
{\cal U}_P&\equiv& \left[2 Q_x Q_y + 2 U_x U_y \right]({\bi \theta})
         \nonumber \\
        &=&(1+\Phi_{xx}+\Phi_{yy})\tl {\cal U}_P+ 2\Phi_{xy} \tl {\cal S}_P,
\label{defin}
\end{eqnarray}
where subscripts $x$ and $y$ stand for respective derivatives in the real
space. 
Equation (\ref{defin}) shows that the measured ${\cal S}_P$, ${\cal Q}_P$ and 
${\cal U}_P$ are
products of the shear tensor and derivatives of the unlensed CMB polarization field. 
Thus the power spectrum of ${\cal S}_P$, ${\cal Q}_P$ and ${\cal U}_P$
will be a convolution of the power in the CMB and that of the
projected mass density.
Convergence and unlensed fields are taken as independent quantities, which 
is a good approximation since most of the lensing power arises from 
low redshifts, which are not correlated with the last scattering surface. 
The general expression for this convolution is
quite involved. 
Here we will discuss it in the limit of large scales relative to
the CMB correlation length $\xi \sim 0.1^{\circ}$ where it is not necessary 
to take the full convolution into
account. In the large scale limit patches of the sky
larger then the correlation length squared are almost independent, which 
simplifies the calculations.
This large scale limit is sufficient to analyze future data from MAP and
Planck satellites and will, in any case, provide a lower limit to the 
attainable signal. 

The reason we have chosen ${\cal S}_P$, ${\cal Q}_P$ and ${\cal U}_P$ 
as variables in consideration is that they have 
transformation properties similar to the Stokes parameters \cite{transfer}.
We wish to show that ${\cal S}_P$, 
${\cal Q}_P$ and ${\cal U}_P$ transform as Stokes parameters during the 
right-handed rotation 
of the spatial base $({\hat{\bi x}},{\hat{\bi y}}, {\hat{\bi n}})$ by an 
angle $\psi$ around ${\hat{\bi n}}$ directed to the observer.
Transformation of Stokes parameters $Q$ and $U$ under rotation are given 
in \cite{spinlong} and their linear combinations $Q \pm i U$ have 
values of spin equal to $\mp 2$,  
\begin{eqnarray}
   Q' \pm i U' & = & e^{\mp i 2\psi} \left(Q \pm i U\right). \label{stokes}
\end{eqnarray}
We can rewrite definitions of polarization  variables (equation \ref{defin}) 
in the following manner
\begin{eqnarray}
  {\cal S}_P &=& (Q+iU)_x(Q-iU)_x + (Q+iU)_y (Q-iU)_y \nonumber \\
  {\cal Q}_P &=& (Q+iU)_x(Q-iU)_x - (Q+iU)_y (Q-iU)_y  \label{trans} \\
  {\cal U}_P &=& (Q+iU)_x(Q-iU)_y + (Q+iU)_y (Q-iU)_x. \nonumber
\end{eqnarray}
When we use equations (\ref{stokes}) and (\ref{trans}) and change variables in
derivatives we find the final expression for transformation of our quantities, 
$$
  \left(\matrix{
        {\cal S}'_P \cr {\cal Q}'_P \cr {\cal U}'_P \cr }\right)
  =
  \left(\matrix{
  1 & 0 & 0 \cr
  0  & \cos2\psi  & \sin2\psi \cr
  0  & -\sin2\psi & \cos2\psi \cr }\right)
  \left(\matrix{
  {\cal S}_P \cr {\cal Q}_P \cr {\cal U}_P \cr }\right).
$$
This demonstrates that ${\cal S}_P$ transforms as a scalar,
while ${\cal Q}_P$ and 
${\cal U}_P$ transform like respective Stokes parameters 
$Q$ and $U$. We will show that these correspond to the familiar 
convergence and shear quantities of gravitational lensing.

In the large scale limit (compared to the correlation length of the CMB 
polarization field) and in the absence  
of lensing we can write the average over the ensemble of the  CMB realizations
\begin{eqnarray}
\langle \tl {\cal S}_P \rangle_{CMB} &=& \sigma_{{\cal S}P} \nonumber \\
\langle \tl {\cal Q}_P \rangle_{CMB} &=& 0 \nonumber \\
\langle \tl {\cal U}_P \rangle_{CMB} &=& 0,  
\end{eqnarray}
where
$\sigma_{{\cal S}P}$  
is defined as  
\begin{eqnarray}
\sigma_{{\cal S}P}=\langle Q_x^2\rangle_{CMB} +\langle Q_y^2\rangle_{CMB}
+\langle U_x^2\rangle_{CMB} +\langle U_y^2\rangle_{CMB}.
\end{eqnarray}
In terms of the polarization power spectrum $C^{\tl P \tl P}_l$ 
it is given by 
\begin{equation}
\sigma_{{\cal S}P} = \int {ldl \over 2\pi}\ l^2 C^{\tl P \tl P}_l, 
\end{equation}
In the presence of lensing the average of 
equation (\ref{defin}) in the large scale limit becomes
\begin{eqnarray}
\langle {\cal S}_P \rangle_{CMB} &=& (1-2\kappa) \sigma_{{\cal S}P} \nonumber \\
\langle {\cal Q}_P \rangle_{CMB} &=& -2 \gamma_1 \sigma_{{\cal S}P} \nonumber \\
\langle {\cal U}_P \rangle_{CMB} &=& -2 \gamma_2 \sigma_{{\cal S}P},
\label{averagecmb}
\end{eqnarray}  
where the shear components $\gamma_1$ and  $\gamma_2$  and convergence $\kappa$ are defined as 
\begin{eqnarray}
\kappa &=& -(\Phi_{xx}+\Phi_{yy})/2 \nonumber\\
\gamma_1 &=& -(\Phi_{xx}-\Phi_{yy})/2 \nonumber\\
\gamma_2 &=& -\Phi_{xy}.
\end{eqnarray}
Physical interpretation of equation (\ref{averagecmb}) is the following.
Convergence $\kappa$ stretches the images and makes the derivatives smaller,
so $\langle {\cal S} \rangle_{CMB}$ is diminished by a factor proportional to $\kappa$.
Similarly, shear produces anisotropy in the derivatives in the same way as in the
images of distinct galaxies.
To normalize the above expressions 
we introduce
\begin{eqnarray}
{\cal S}_P^\prime&=&-{{\cal S}_P\over \sigma_{{\cal S}P}}+1 \nonumber \\
{\cal Q}_P^\prime&=&-{{\cal Q}_P\over \sigma_{{\cal S}P}} \nonumber \\
{\cal U}_P^\prime&=&-{{\cal U}_P\over \sigma_{{\cal S}P}},
\label{redef}
\end{eqnarray}
such that
\begin{eqnarray}
\langle {\cal S}_P^{\prime} \rangle_{CMB} &=& 2\kappa\nonumber \\
\langle {\cal Q}_P^{\prime} \rangle_{CMB} &=& 2 \gamma_1 \nonumber \\
\langle {\cal U}_P^{\prime} \rangle_{CMB} &=& 2 \gamma_2.
\label{avgprime}
\end{eqnarray}
In the following we will only
use these variables so we drop the primes from now on. 

Instead of working with rotationally non-invariant quantities ${\cal Q}_P$ and ${\cal U}_P$
we combine them to form scalar field ${\cal E}_P$ and 
pseudo-scalar ${\cal B}_P$. In the Fourier space these have the form
\begin{eqnarray}
{\cal E}_P({\bi l})
         &=&  {\cal Q}_P({\bi l} ) \cos(2\phi_{\bi l})+
         {\cal U}_P({\bi l} ) \sin(2\phi_{\bi l}) \nonumber \\
{\cal B}_P({\bi l})
         &=&  {\cal Q}_P({\bi l}) \sin(2\phi_{\bi l})-
         {\cal U}_P({\bi l} ) \cos(2\phi_{\bi l}), 
\label{scalars}
\end{eqnarray}
where $\phi_{\bi l}$ is the azimuthal angle of the mode ${\bi l}$.
In the Fourier space convergence and shear are related to each other through the
relations  
\begin{equation}
\gamma_1(\bi{l})=\kappa(\bi{l})\cos(2\phi_l)\;\;\;\;\;\;
\gamma_2(\bi{l})=\kappa(\bi{l})\sin(2\phi_l),
\label{kappaft}
\end{equation}
 From equations (\ref{avgprime}), (\ref{scalars}) and (\ref{kappaft}) it
follows that
\begin{eqnarray}
\langle {\cal S}_P \rangle_{CMB} &=& 2\kappa \nonumber \\
\langle {\cal E}_P \rangle_{CMB} &=& 2\kappa \nonumber \\
\langle {\cal B}_P \rangle_{CMB} &=&0.
\label{recon}
\end{eqnarray}
Thus the convergence $\kappa$ can be reconstructed in two ways, either from 
${\cal S}_P$ or ${\cal E}_P$. Moreover, vanishing of $ {\cal B}_P $ 
on average can be helpful in indentifing the  
cosmological part  of the signal or removing some foregrounds or other systematics.

To reconstruct the convergence power spectrum $C_l^{\kappa \kappa}$ 
we form
its estimators     
\begin{equation}
\hat{C}^{\cal WW'}_l= {1 \over 2} \left[ {\cal W}({\bi l})^*{\cal W}'({\bi l'})+
{\cal W}'({\bi l'})^*{\cal W}({\bi l})\right] \delta_{ll'},
\end{equation}
where ${\cal W}$ or ${\cal W}'$ stand for ${\cal S}_P$, ${\cal E}_P$,
${\cal S}_T$ or ${\cal E}_T$, with the first two obtained from polarization 
and the last two 
from temperature \cite{lens}. 
One can write the mean value of the estimator as 
\begin{equation}
\langle \hat{C}_l^{{\cal WW}'} \rangle_{CMB} = 4C_l^{\kappa \kappa}+N_l^{\cal WW'}.
\label{estim}
\end{equation}
Here $N_l^{\cal WW'}$ is a power spectrum of the noise arising from the
intrinsic fluctuations in 
CMB temperature or polarization (or both for cross-correlation).
This noise arises from the random
nature of the CMB and has to be carefully examined to be properly subtracted 
from the estimated power spectrum $\hat{C}^{\cal WW'}_l$ to get an unbiased 
convergence spectrum $C_l^{\kappa \kappa}$.

\begin{figure}[p]
\begin{center}
\vspace*{10cm}
\includegraphics{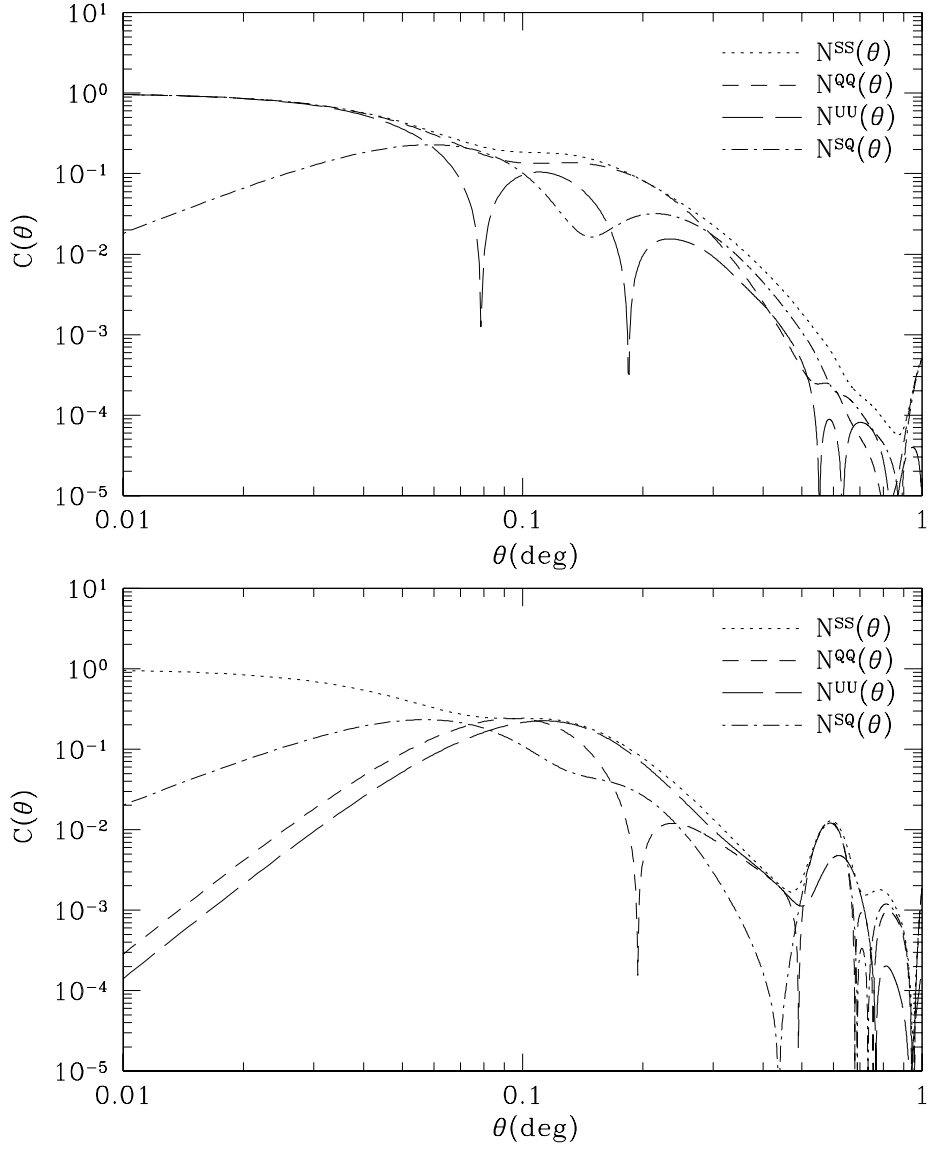}
\end{center}
\caption{The upper panel shows auto-correlation functions for polarization  
$N^{{\cal S}_P {\cal S}_P}(\theta)$, $N^{{\cal Q}_P {\cal Q}_P}(\theta)$,
$N^{{\cal U}_P {\cal U}_P}(\theta)$ and $N^{{\cal S}_P {\cal Q}_P}(\theta)$.
The lower panel shows the cross-correlations  
$N^{{\cal S}_T {\cal S}_P}(\theta)$, $N^{{\cal Q}_T {\cal Q}_P}(\theta)$,
$N^{{\cal U}_T {\cal U}_P}(\theta)$ and $N^{{\cal S}_T {\cal Q}_P}(\theta)$.
Both plots are for the cosmological concordance model (see text) and
without detector noise.}
\label{figcorr}
\end{figure}

The noise power spectra can be computed analytically. First, we consider 
correlation functions between all the quantities  ${\tl {\cal S}}_P$,
${\tl {\cal Q}}_P$, ${\tl {\cal U}}_P$, 
${\tl {\cal S}}_T$, 
${\tl {\cal Q}}_T$ and ${\tl {\cal U}}_T$. They are all defined as 
combinations of derivatives so using 
equations (\ref{defin}) and (\ref{redef}) 
results in the following expressions for the noise 
\begin{eqnarray}
N^{{\cal S}_P {\cal S}_P}(\theta) &\equiv &
\langle {\tl {\cal S}}_P(0) {\tl {\cal S}}_P(\theta)\rangle_{CMB} \nonumber \\
&=& {2 \over \sigma_{{\cal S}P}^2 }
\left[ (C_{xx}^{QQ})^2+(C_{yy}^{QQ})^2 + (C_{xx}^{UU})^2+(C_{yy}^{UU})^2+
4 (C_{xy}^{QU})^2 \right] \nonumber \\
N^{{\cal Q}_P {\cal Q}_P}(\theta) &\equiv &
\langle {\tl {\cal Q}}_P(0) {\tl {\cal Q}}_P(\theta)\rangle_{CMB} \nonumber \\
&=& {2 \over \sigma_{{\cal S}P}^2 }
\left[ (C_{xx}^{QQ})^2+(C_{yy}^{QQ})^2 + (C_{xx}^{UU})^2+(C_{yy}^{UU})^2 -
4 (C_{xy}^{QU})^2 \right] \nonumber \\
N^{{\cal U}_P {\cal U}_P}(\theta) &\equiv &
\langle {\tl {\cal U}}_P(0) {\tl {\cal U}}_P(\theta)\rangle_{CMB} \nonumber \\
&=& {4 \over \sigma_{{\cal S}P}^2 }
\left[ C_{xx}^{QQ}\ C_{yy}^{QQ} + C_{xx}^{UU}\ C_{yy}^{UU} +
2(C_{xy}^{QU})^2 \right] \nonumber\\
N^{{\cal S}_P {\cal Q}_P}(\theta) &\equiv &
\langle {\tl {\cal S}}_P(0) {\tl {\cal Q}}_P(\theta)\rangle_{CMB} \nonumber \\
&=& {2 \over \sigma_{{\cal S}P}^2 }
\left[ (C_{xx}^{QQ})^2 - (C_{yy}^{QQ})^2 + (C_{xx}^{UU})^2 - (C_{yy}^{UU})^2
\right] \nonumber\\
N^{{\cal S}_T {\cal S}_P}(\theta) &\equiv &
\langle {\tl {\cal S}}_T(0) {\tl {\cal S}}_P(\theta)\rangle_{CMB} \nonumber \\
&=& {2 \over \sigma_{{\cal S}T} \sigma_{{\cal S}P} }
\left[ (C_{xx}^{TQ})^2+(C_{yy}^{TQ})^2 + 2 (C_{xy}^{TU})^2 \right]  \nonumber\\
N^{{\cal Q}_T {\cal Q}_P}(\theta) &\equiv &
\langle {\tl {\cal Q}}_T(0) {\tl {\cal Q}}_P(\theta)\rangle_{CMB} \nonumber \\
&=& {2 \over \sigma_{{\cal S}T} \sigma_{{\cal S}P} }
\left[ (C_{xx}^{TQ})^2+(C_{yy}^{TQ})^2 - 2 (C_{xy}^{TU})^2 \right] \nonumber \\
N^{{\cal U}_T {\cal U}_P}(\theta) &\equiv &
\langle {\tl {\cal U}}_T(0) {\tl {\cal U}}_P(\theta)\rangle_{CMB} \nonumber \\
&=& {4 \over \sigma_{{\cal S}T} \sigma_{{\cal S}P} }
\left[ C_{xx}^{TQ}\ C_{yy}^{TQ} + (C_{xy}^{TU})^2 \right] \nonumber\\
N^{{\cal S}_T {\cal Q}_P}(\theta) &\equiv &
\langle {\tl {\cal S}}_T(0) {\tl {\cal Q}}_P(\theta)\rangle_{CMB} \nonumber \\
&=& {2 \over \sigma_{{\cal S}T} \sigma_{{\cal S}P} }
\left[ (C_{xx}^{TQ})^2 - (C_{yy}^{TQ})^2 \right].
\label{correl}
\end{eqnarray}
Those for temperature are given in \cite{lens} and will not be repeated here.
The correlation functions $C^{XY}_{xy}$ are given in the Appendix.
Equations (\ref{correl}) together with equations 
(\ref{appform}) contain all the necessary 
information for noise estimate. 
In figure \ \ref{figcorr} we show correlation functions for polarization
and cross-correlation between temperature and polarization. We assume 
throughout that the underlying model
is the concordance model \cite{ostriker} with $\Omega_b=0.04$,
$\Omega_m=0.33$,
$\Omega_{\Lambda}=0.63$, $h=0.65$ and $n_s=1$. As discussed in \cite{zs98}
other viable models give comparable results for signal to noise. 
For this model the 
correlation length for polarization is $\xi_{PP} \sim 0.07^{\circ}$, while
for cross-correlation
$\xi_{TP} \sim 0.1^{\circ}$ and for temperature $\xi_{TT} \sim 0.15^{\circ}$. 
Each patch of size $\xi^2$ is roughly independent and decreasing $\xi$ 
by a factor of 2 therefore increases by 4 number of independent patches. 
Since on large scales CMB noise behaves roughly as a white noise this 
reduces the noise level by corresponding factor.
This implies that if we could observe polarization with perfect resolution 
it would give significantly larger signal to noise than temperature alone.
However, since polarization has a weaker signal than temperature this 
advantage becomes possible only for very sensitive detectors and small 
beams, as discussed below in more detail.

\begin{figure}[p]
\begin{center}
\vspace*{10cm}
\includegraphics{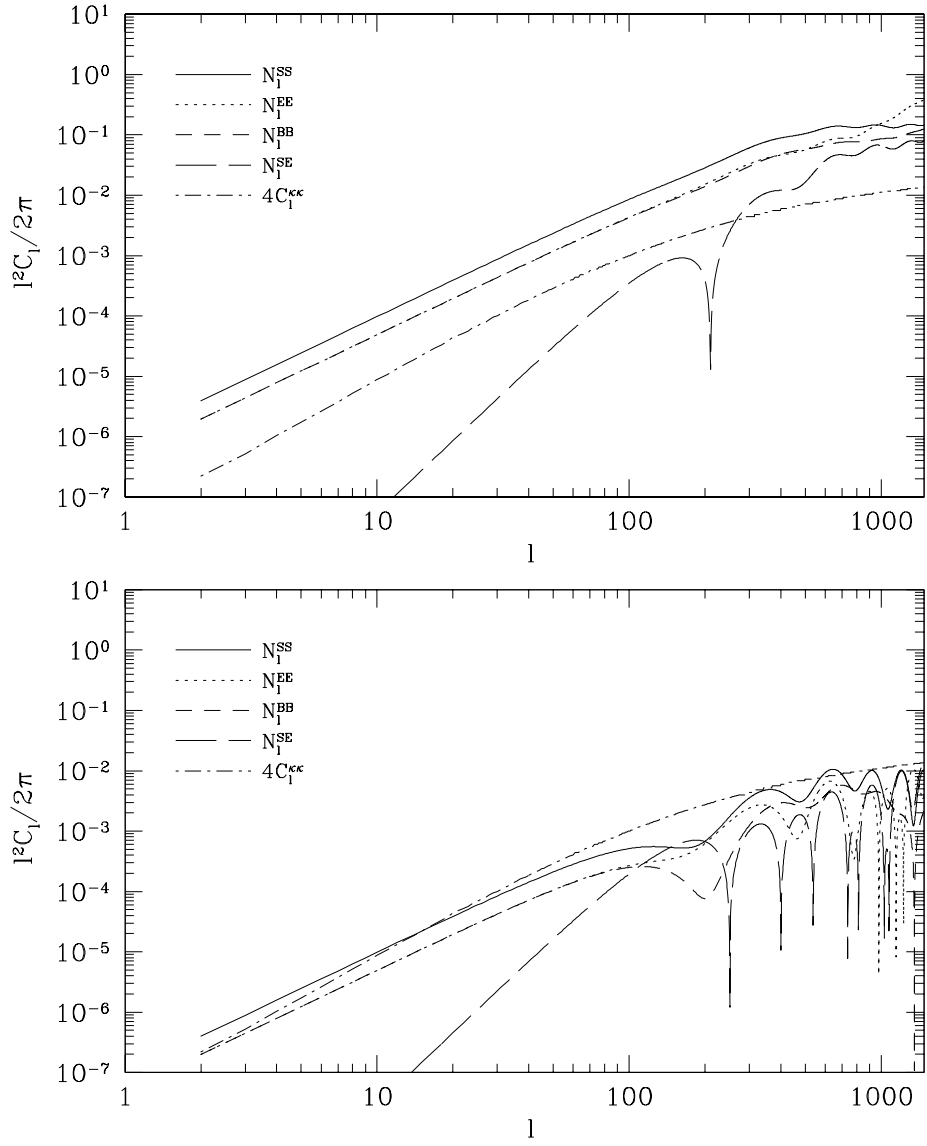}
\end{center}
\caption{
In the upper panel there are noise power spectra for polarization 
$N^{{\cal S}_P {\cal S}_P}_l$, $N^{{\cal E}_P {\cal E}_P}_l$, 
$N^{{\cal B}_P {\cal B}_P}_l$ and $N^{{\cal S}_P {\cal E}_P}_l$,
yet in the lower panel for the cross-correlation
$N^{{\cal S}_T {\cal S}_P}_l$, $N^{{\cal E}_T {\cal E}_P}_l$,
$N^{{\cal B}_T {\cal B}_P}_l$ and $N^{{\cal S}_T {\cal E}_P}_l$.
They are coupled with correlation functions presented in fig.\ref{figcorr} 
via equations (\ref{defcl}). }
\label{figpower}
\end{figure}

The noise power spectra can be obtained from the real space
correlations using the following expressions \cite{lensps}   
\begin{eqnarray}
N^{{\cal S}_A {\cal S}_B}_l&=&2\pi \int  \theta d\theta \ 
N^{{\cal S}_A {\cal S}_B}(\theta) \ J_0(l\theta) \nonumber \\
N^{{\cal E}_A {\cal E}_B}_l&=&\pi \int  \theta d\theta
\left\{ \left[N^{{\cal Q}_A {\cal Q}_B}(\theta)+N^{{\cal U}_A {\cal U}_B}(\theta)\right] \ J_0(l\theta)
+ \left[N^{{\cal Q}_A {\cal Q}_B}(\theta)-N^{{\cal U}_A {\cal U}_B}(\theta)\right]\ J_4(l\theta)\right\}  \nonumber \\
N^{{\cal B}_A {\cal B}_B}_l&=&\pi \int  \theta d\theta \
\left\{ \left[N^{{\cal Q}_A {\cal Q}_B}(\theta)+N^{{\cal U}_A {\cal U}_B}(\theta)\right]\ J_0(l\theta)
- \left[N^{{\cal Q}_A {\cal Q}_B}(\theta)-N^{{\cal U}_A {\cal U}_B}(\theta)\right]\ J_4(l\theta)\right\}  \nonumber \\
N^{{\cal S}_A {\cal E}_B}_l&=&2\pi \int  \theta d\theta \
N^{{\cal S}_A {\cal Q}_B}(\theta)\ J_2(l\theta),
\label{defcl}
\end{eqnarray}
where $A$ and $B$ stand for $T$ or $P$.
%
%
 From this one obtains the CMB noise power spectra 
\begin{eqnarray}
N^{{\cal S}_P {\cal S}_P}_l &=& {\pi \over 2\sigma_{{\cal S}P}^2}
\int \theta d\theta \left\{
     2(C_0^{\tl P \tl P})^2+ 3(C_2^{\tl P \tl P})^2 + 2(C_4^{\tl P \tl P})^2
     + (C_6^{\tl P \tl P})^2 \right\}\ J_0(l \theta) \nonumber \\
N^{{\cal E}_P {\cal E}_P}_l &=& {\pi \over \sigma_{{\cal S}P}^2}
\int \theta d\theta \left\{
     (C_0^{\tl P \tl P})^2 + (C_4^{\tl P \tl P})^2 
     \right\}\ J_0(l \theta) + \left\{
     (C_2^{\tl P \tl P})^2 +
     2 C_2^{\tl P \tl P} C_6^{\tl P \tl P} \right\}\ J_4(l \theta) \nonumber \\
N^{{\cal B}_P {\cal B}_P}_l &=& {\pi \over \sigma_{{\cal S}P}^2}
\int \theta d\theta \left\{
     (C_0^{\tl P \tl P})^2 + (C_4^{\tl P \tl P})^2 
     \right\}\ J_0(l \theta) - \left\{
     (C_2^{\tl P \tl P})^2 +
     2 C_2^{\tl P \tl P} C_6^{\tl P \tl P} \right\}\ J_4(l \theta) \nonumber\\
N^{{\cal S}_P {\cal E}_P}_l &=& - {\pi \over \sigma_{{\cal S}P}^2}
\int \theta d\theta \left\{
     2 C_0^{\tl P \tl P} C_2^{\tl P \tl P} +
     C_2^{\tl P \tl P} C_4^{\tl P \tl P} +
     C_4^{\tl P \tl P} C_6^{\tl P \tl P} \right\}\ J_2(l \theta) \nonumber\\
N^{{\cal S}_T {\cal S}_P}_l &=&
{\pi \over \sigma_{{\cal S}T} \sigma_{{\cal S}P} }
\int \theta d\theta \left\{
     (C_0^{\tl T \tl P})^2+ 2(C_2^{\tl T \tl P})^2 + 
     4(C_4^{\tl T \tl P})^2 \right\}\ J_0(l \theta) \nonumber\\
N^{{\cal E}_T {\cal E}_P}_l &=&
{2\pi \over \sigma_{{\cal S}T} \sigma_{{\cal S}P}}
\int \theta d\theta \left\{
     (C_2^{\tl T \tl P})^2 \ J_0(l \theta) +
     C_0^{\tl T \tl P} C_4^{\tl T \tl P} \ J_4(l \theta) \right\}\nonumber \\
N^{{\cal B}_T {\cal B}_P}_l &=&
{2\pi \over \sigma_{{\cal S}T} \sigma_{{\cal S}P}}
\int \theta d\theta \left\{
     (C_2^{\tl T \tl P})^2 \ J_0(l \theta) -
     C_0^{\tl T \tl P} C_4^{\tl T \tl P} \ J_4(l \theta) \right\} \nonumber\\
N^{{\cal S}_T {\cal E}_P}_l &=&
-{2\pi \over \sigma_{{\cal S}T} \sigma_{{\cal S}P}}
\int \theta d\theta \left\{
     C_0^{\tl T \tl P} + C_4^{\tl T \tl P} \right\}\ 
     C_2^{\tl T \tl P} \ J_2(l \theta)
\label{noisespec} 
\end{eqnarray}
We  do not need to consider spectra containing ${\cal B}_T$ or ${\cal B}_P$ 
as these pseudo-scalar quantities do not correlate with scalars. 
Fig.\ \ref{figpower} shows noise power spectra for 
polarization and cross-correlation.  The most visible feature is the white noise
behavior on large scales, which
confirms that derivatives of temperature and
Stokes parameters are almost 
uncorrelated on scales above a fraction of a degree.

Let us consider now the power spectra in the limit of low $l$. 
Utilizing of the Bessel functions properties in this limit 
and their 
orthonormality relations we obtain from equation (\ref{noisespec}) the
following limits 
\begin{eqnarray}
\lim_{l\rightarrow 0} N_l^{{\cal S}_P {\cal S}_P} &=&
4\pi {\int l^5dl (C_l^{\tl P \tl P})^2 \over (\int l^3 dlC^{\tl P \tl P}_l)^2} \nonumber\\
\lim_{l\rightarrow 0} N_l^{{\cal E}_P {\cal E}_P} &=&
\lim_{l\rightarrow 0} N_l^{{\cal B}_P {\cal B}_P} =
{1 \over 2} \lim_{l\rightarrow 0} N_l^{{\cal S}_P {\cal S}_P} \label{limitp}\\
\lim_{l\rightarrow 0} N_l^{{\cal S}_P {\cal E}_P} &=& 0 \nonumber\\
\lim_{l\rightarrow 0} N_l^{{\cal S}_T {\cal S}_P} &=&
4\pi {\int l^5dl (C_l^{\tl T \tl P})^2 \over
\int l^3 dlC^{\tl T \tl T}_l \ \int l^3 dlC^{\tl P \tl P}_l } \nonumber\\
\lim_{l\rightarrow 0} N_l^{{\cal E}_T {\cal E}_P} &=&
\lim_{l\rightarrow 0} N_l^{{\cal B}_T {\cal B}_P} =
{1 \over 2} \lim_{l\rightarrow 0} N_l^{{\cal S}_T {\cal S}_P} \label{limitc}\\
\lim_{l\rightarrow 0} N_l^{{\cal S}_T {\cal E}_P} &=& 0. \nonumber
\end{eqnarray}
The power spectra presented in the fig.\ \ref{figpower} confirm this large 
scale behavior. It is worth noting that
the noise for ${\cal S}$ variables is two times higher than for ${\cal E}$ and ${\cal B}$, although
normalized variances of ${\tl {\cal S}}_P$, ${\tl {\cal Q}}_P$ and ${\tl {\cal U}}_P$ are all the same.
This is because of the spin nature of ${\tl {\cal Q}}_P$ and  ${\tl {\cal U}}_P$.
The implication of this is that 
the reconstructed convergence power spectrum will have the noise amplitude
two times lower when we use ${\cal E}_P$ instead of ${\cal S}_P$.

When we have quantified the intrinsic CMB noise we can use
equation
(\ref{estim}) to obtain the best estimate for
$\hat{C}_l^{\kappa \kappa} $ 
from observations. In addition, we also need to determine the error on 
the estimated power spectrum
$\hat{C}^{\cal WW'}_l$. To proceed analytically 
we make an additional assumption that Fourier transforms of 
quantities ${\cal S}_P$, ${\cal E}_P$,
${\cal S}_T$ and ${\cal E}_T$ are Gaussian distributed. From 
simulations presented in \cite{lens} this seems to be a good approximation
on large scales, where a single long wavelength mode receives 
contribution from 
many almost independent structures in real space
and the central limit theorem makes them almost Gaussian.    
In general, from ${\cal S}_T$, ${\cal E}_T$, ${\cal S}_P$ and ${\cal E}_P$ we can construct 
10 estimators of $C_l^{\kappa \kappa}$, 
3 for the temperature field, 3 for polarization and 4 for cross-correlation. 
Under these assumptions we obtain the elements of the covariance matrices of 
the polarization and 
cross-correlation estimators
\begin{eqnarray}
{\rm Var}\left(\hat{C}^{{\cal S}_A {\cal S}_B}_l\right) &=&
{1\over 2l+1} \left[(4C_l^{\kappa \kappa} + N^{{\cal S}_A {\cal S}_B}_l)^2 +
(4C_l^{\kappa \kappa}+N^{{\cal S}_A{\cal S}_A}_l)(4C_l^{\kappa \kappa}+ N^{{\cal S}_B{\cal S}_B}_l)\right] \\
{\rm Var}\left(\hat{C}^{{\cal E}_A {\cal E}_B}_l\right) &=&
{1\over 2l+1} \left[(4C_l^{\kappa \kappa}+N^{{\cal E}_A {\cal E}_B}_l)^2 +
(4C_l^{\kappa \kappa}+N^{{\cal E}_A{\cal E}_A}_l)(4C_l^{\kappa \kappa}+ N^{{\cal E}_B{\cal E}_B}_l)\right] \\
{\rm Var}\left(\hat{C}^{{\cal B}_A {\cal B}_B}_l\right) &=&
{1\over 2l+1}\left[(N^{{\cal B}_A {\cal B}_B}_l)^2 +
N^{{\cal B}_A{\cal B}_A}_l N^{{\cal B}_B{\cal B}_B}_l\right] \\
{\rm Var}\left(\hat{C}^{{\cal S}_A {\cal E}_B}_l\right) &=&
{1\over 2l+1}\left[(4C_l^{\kappa \kappa}+N^{{\cal S}_A {\cal E}_B}_l)^2 +
(4C_l^{\kappa \kappa}+N^{{\cal S}_A{\cal S}_A}_l)(4C_l^{\kappa \kappa}+ N^{{\cal E}_B{\cal E}_B}_l)\right] \\
{\rm Var}\left(\hat{C}^{{\cal E}_A {\cal S}_B}_l\right) &=&
{1\over 2l+1}\left[(4C_l^{\kappa \kappa}+N^{{\cal E}_A {\cal S}_B}_l)^2 +
(4C_l^{\kappa \kappa}+N^{{\cal E}_A{\cal E}_A}_l)(4C_l^{\kappa \kappa}+ N^{{\cal S}_B{\cal S}_B}_l)\right] \\
{\rm Cov}\left(\hat{C}^{{\cal S}_A {\cal S}_B}_l, \hat{C}^{{\cal E}_A {\cal E}_B}_l\right) &=&
{1\over 2l+1}\left[(4C_l^{\kappa \kappa}+N^{{\cal S}_A {\cal E}_B}_l)^2 +
(4C_l^{\kappa \kappa}+N^{{\cal S}_A{\cal E}_A}_l)(4C_l^{\kappa \kappa}+ N^{{\cal S}_B{\cal E}_B}_l)\right] \\
{\rm Cov}\left(\hat{C}^{{\cal S}_A {\cal S}_B}_l, \hat{C}^{{\cal S}_A {\cal E}_B}_l\right) &=&
{1\over 2l+1}\left[(4C_l^{\kappa \kappa}+N^{{\cal S}_A {\cal S}_B}_l)
(4C_l^{\kappa \kappa}+N^{{\cal S}_A {\cal E}_B}_l) +
(4C_l^{\kappa \kappa}+N^{{\cal S}_A{\cal S}_A}_l)(4C_l^{\kappa \kappa}+ N^{{\cal S}_B{\cal E}_B}_l)\right] \\
{\rm Cov}\left(\hat{C}^{{\cal E}_A {\cal E}_B}_l, \hat{C}^{{\cal S}_A {\cal E}_B}_l\right) &=&
{1\over 2l+1}\left[(4C_l^{\kappa \kappa}+N^{{\cal E}_A {\cal E}_B}_l)
(4C_l^{\kappa \kappa}+N^{{\cal S}_A {\cal E}_B}_l) +
(4C_l^{\kappa \kappa}+N^{{\cal S}_A{\cal E}_A}_l)(4C_l^{\kappa \kappa}+ N^{{\cal E}_B{\cal E}_B}_l)\right] \\
{\rm Cov}\left(\hat{C}^{{\cal E}_A {\cal S}_B}_l, \hat{C}^{{\cal S}_A {\cal E}_B}_l\right) &=&
{1\over 2l+1}\left[(4C_l^{\kappa \kappa}+ N^{{\cal E}_A {\cal E}_B}_l)
(4C_l^{\kappa \kappa}+N^{{\cal S}_A {\cal S}_B}_l) +
(4C_l^{\kappa \kappa}+N^{{\cal S}_A{\cal E}_A}_l)(4C_l^{\kappa \kappa}+N^{{\cal S}_B{\cal E}_B}_l)\right] \\
{\rm Cov}\left(\hat{C}^{{\cal E}_A {\cal S}_B}_l, \hat{C}^{{\cal S}_A {\cal S}_B}_l\right) &=&
{1\over 2l+1}\left[(4C_l^{\kappa \kappa}+ N^{{\cal E}_A {\cal S}_A}_l)
(4C_l^{\kappa \kappa}+N^{{\cal S}_B {\cal S}_B}_l) +
(4C_l^{\kappa \kappa}+N^{{\cal E}_A{\cal S}_B}_l)(4C_l^{\kappa \kappa}+ N^{{\cal S}_A{\cal S}_B}_l)\right] \\
{\rm Cov}\left(\hat{C}^{{\cal E}_A {\cal S}_B}_l, \hat{C}^{{\cal E}_A {\cal E}_B}_l\right) &=&
{1\over 2l+1}\left[(4C_l^{\kappa \kappa}+N^{{\cal E}_A {\cal E}_B}_l)
(4C_l^{\kappa \kappa}+N^{{\cal S}_B {\cal E}_A}_l) +
(4C_l^{\kappa \kappa}+N^{{\cal E}_A{\cal E}_A}_l)(4C_l^{\kappa \kappa}+ N^{{\cal S}_B{\cal E}_B}_l)\right] 
\end{eqnarray}
where $A$ and $B$ stand for $T$ or $P$ again. 

To assess the variance of the overall estimator $\hat{C}_l^{\kappa \kappa}$
that is a combination of three estimators in the case of polarization and
four in the case of cross-correlation (eq. \ref{estim}) we
treat our variables
${\cal S}_T$, ${\cal S}_P$, ${\cal E}_T$ and ${\cal E}_P$ as Gaussian.
Using the Fisher information matrix we can derive the desired
variances \cite{maxt}
\begin{eqnarray}
{\cal F}_{ij} ={1\over 2} {\rm Tr} 
\left[{\rm Cov}^{-1}({\cal W},{\cal W'}){\partial {\rm Cov}({\cal W},{\cal W'}) \over \partial s_i}
{\rm Cov}^{-1}({\cal W},{\cal W'}){\partial {\rm Cov}({\cal W},{\cal W'}) \over \partial s_j}\right].
\end{eqnarray}
The inverse of the Fisher matrix
gives the covariance matrix.
The Fisher matrix is one dimensional and relevant
derivatives are taken with respect to $C_l^{\kappa \kappa}$.
We find the following expression  for variance
of the $\hat{C}_l^{\kappa \kappa}$ estimator when we take into account
only polarization information
\begin{eqnarray}
{\rm Var} \left(\hat{C}_l^{\kappa \kappa}\right)_{PP} =
{1\over 8(2l+1)} {\left[ C^{{\cal S}_P{\cal S}_P}_l C^{{\cal E}_P {\cal E}_P}_l
- \left(C^{{\cal S}_P{\cal E}_P}_l\right)^2\right]^2 \over
\left[C^{{\cal S}_P{\cal S}_P}_l + C^{{\cal E}_P {\cal E}_P}_l
- 2 C^{{\cal S}_P{\cal E}_P}_l\right]^2}
\label{generp}
\end{eqnarray}
Similarly, using the cross-correlation variables only
the variance is
\begin{eqnarray}
{\rm Var} \left(\hat{C}_l^{\kappa \kappa}\right)_{TP} =
{1\over 8(2l+1)}{
\left[C^{{\cal S}_T{\cal S}_T}_l C^{{\cal E}_T {\cal E}_T}_l
- \left(C^{{\cal S}_T{\cal E}_T}_l\right)^2\right]
\left[C^{{\cal S}_P{\cal S}_P}_l C^{{\cal E}_P {\cal E}_P}_l
- \left(C^{{\cal S}_P{\cal E}_P}_l\right)^2\right]
\over
\left[C^{{\cal S}_T{\cal S}_T}_l + C^{{\cal E}_T {\cal E}_T}_l
- 2 C^{{\cal S}_T{\cal E}_T}_l\right]^2
\left[C^{{\cal S}_P{\cal S}_P}_l + C^{{\cal E}_P {\cal E}_P}_l
- 2 C^{{\cal S}_P{\cal E}_P}_l\right]^2},
\label{genertp}
\end{eqnarray}
where we have dropped the cross-correlation terms    
$C^{{\cal W}_T{\cal W}_P}$ which are much smaller than the diagonal terms 
in the large scale limit. For an experiment like MAP or Planck we will have  
$N^{{\cal S}_P{\cal S}_P} \gg N^{{\cal S}_T{\cal S}_T} \gg
N^{{\cal S}_T{\cal S}_P}$ and the same for other combinations of ${\cal S}_T$, 
${\cal E}_T$, ${\cal S}_P$, ${\cal E}_P$, as can be seen from the noise spectra 
in figure \ref{power}. Only for a significantly more sensitive detector 
will the three noise spectra become comparable in which case, as argued above, 
polarization noise power spectrum will become smaller than that of  the
temperature. For this reason we provide separately the information from 
polarization and from temperature polarization cross-correlation.
In addition,
the cosmic variance term proportional to $4C_l^{\kappa \kappa}$ is also
much smaller than the CMB intrinsic noise contribution and can be dropped.
The main source of errors is then the intrinsic CMB noise.
In the large scale limit
$N^{{\cal S}_P{\cal E}_P}$ is more than an order of magnitude less than
$N^{{\cal S}_P{\cal S}_P}$, $N^{{\cal E}_P{\cal E}_P}$ or
$N^{{\cal B}_P{\cal B}_P}$ that implies we
can neglect that term in the covariances.
These approximations lead to a diagonal form of covariance matrices
that can be written as 
\begin{eqnarray}
{\rm Var} \left(\hat{C}_l^{\kappa \kappa}\right)_{PP} &=&
8(2l+1) \left[
{1 \over (C^{{\cal S}_T{\cal S}_T}_l)^2} +
{1 \over (C^{{\cal E}_T{\cal E}_T}_l)^2} +
{2 \over C^{{\cal S}_T{\cal S}_T}_l C^{{\cal E}_T{\cal E}_T}_l}
\right]  \\ 
{\rm Var} \left(\hat{C}_l^{\kappa \kappa}\right)_{TP} &=&
8(2l+1) \left[
{1 \over C^{{\cal S}_T{\cal S}_T}_l C^{{\cal S}_P{\cal S}_P}_l}+
{1 \over C^{{\cal E}_T{\cal E}_T}_l C^{{\cal E}_P{\cal E}_P}_l}+
{1 \over C^{{\cal S}_T{\cal S}_T}_l C^{{\cal E}_P{\cal E}_P}_l}+
{1 \over C^{{\cal E}_T{\cal E}_T}_l C^{{\cal S}_P{\cal S}_P}_l}
\right]
\label{covdiagcor}
\end{eqnarray}
The minimum variance combination of these estimators in 
the large scale limit is 
for polarization
\begin{eqnarray}
\langle 4\hat{C}^{\kappa \kappa}_l \rangle =
{1 \over 9}\left(\langle \hat{C}_l^{{\cal S}_P {\cal S}_P} \rangle -N_l^{{\cal S}_P {\cal S}_P}\right)+
{4 \over 9}\left(\langle \hat{C}_l^{{\cal E}_P {\cal E}_P} \rangle-N_l^{{\cal E}_P {\cal E}_P}\right)
+ {4 \over 9}\left(\langle \hat{C}_l^{{\cal S}_P {\cal E}_P} \rangle-N_l^{{\cal S}_P {\cal E}_P}\right)
\label{ckkpp}
\end{eqnarray}
and for cross-correlation
\begin{eqnarray}
\langle 4\hat{C}^{\kappa \kappa}_l\rangle =
{1 \over 9}\left(\langle \hat{C}_l^{{\cal S}_T {\cal S}_P}\rangle-N_l^{{\cal S}_T {\cal S}_P}\right)+
{4 \over 9}\left(\langle \hat{C}_l^{{\cal E}_T {\cal E}_P}\rangle-N_l^{{\cal E}_T {\cal E}_P}\right) 
+ {2 \over 9}\left(\langle \hat{C}_l^{{\cal S}_T {\cal E}_P}\rangle-N_l^{{\cal S}_T {\cal E}_P}\right) +
{2 \over 9}\left(\langle \hat{C}_l^{{\cal E}_T {\cal S}_P}\rangle -N_l^{{\cal E}_T {\cal S}_P}\right)
\label{ckktp}
\end{eqnarray}
In the limit $l \rightarrow 0$ variances of these overall estimators
can be written as
\begin{eqnarray}
{\rm Var} \left(4\hat{C}_l^{\kappa \kappa}\right)_{PP} &=&
{2\over 9(2l+1)}(N^{{\cal S}_P {\cal S}_P}_l)^2
\label{varpp} \\
{\rm Var} \left(4\hat{C}_l^{\kappa \kappa}\right)_{TP} &=&
{1\over 9(2l+1)} N^{{\cal S}_T {\cal S}_T}_l
N^{{\cal S}_P {\cal S}_P}_l
\label{vartp}
\end{eqnarray}
These expressions show that if $N^{{\cal S}_T {\cal S}_T}_l \ll 
N^{{\cal S}_P {\cal S}_P}_l$ then it will be the cross-correlation 
that will add most of polarization information to $C_l^{\kappa \kappa}$.
This information increase will be small compared 
to the information obtained from the temperature alone, but may still 
provide important independent confirmation. If 
$N^{{\cal S}_T {\cal S}_T}_l \sim N^{{\cal S}_P {\cal S}_P}_l$
polarization provides as much information to $C_l^{\kappa \kappa}$ 
as temperature.
In further considerations we will 
use the complete expressions for respective variances without the 
above approximations. 

Another consequence of the intrinsic CMB fluctuations is that
we cannot recover the convergence 
from the individual structures, because the coherence length of CMB 
is too large. We will show that it is still
possible to measure the convergence in a statistical sense, by 
averaging over many Fourier modes to extract the power spectrum.
We will determine the signal to noise for various detector 
sensitivities in the following section.

\section{Convergence from observations}
\label{sec2}

Before computing the expected signal to noise for future missions we
need to include the effect of detector noise and angular resolution
on extracting the convergence power spectrum. Both of 
these have been discussed in \cite{lens} and the expressions 
derived there can be applied 
to polarization as well.

To quantify the influence of the detector beam on measured
Stokes parameters one introduces a filter function $F({\bi \theta})$, which
describes the detector beam filtering of the data. The
Stokes parameters are convolved with this function
\begin{eqnarray}
X_a({\bi \theta})=\int F(\bi{\theta}-\bi{\theta}')\ X_a(\bi{\theta}')
d^2{\bi \theta}' \label{convo}
\end{eqnarray}
where $X$ stands for Stokes parameters $Q$ or $U$.
Using equation (\ref{dert}) and Fourier decomposition equation
(\ref{convo}) becomes
\begin{eqnarray}
X_a({\bi \theta})={1 \over (2\pi)^{2}} \int d^2{\bi l} F(l) X_a({\bi l})
       e^{i{\bi l}\cdot \bi{\theta}}
+{1 \over (2\pi)^{4}} \int d^2{\bi l} \int d^2{\bi q} F(|\bi l+ \bi q|)
X_b({\bi l}) \Phi_{ab} (\bi q) e^{i{(\bi l+ \bi q)}\cdot \bi{\theta}}.
\label{filter1}
\end{eqnarray}
In the presence of filter $F({\bi \theta})$ it follows from 
equations (\ref{defin}) and (\ref{filter1})
\begin{eqnarray}
\langle {\cal S}_P({\bi \theta}) \rangle_{CMB} =
{1 \over (2\pi)^{2}} \int d^2{\bi l} F^2(l) l^2 C^{\tl P \tl P}_l
\left[ 1 - {1 \over (2\pi)^{2}} \int d^2{\bi q} \ 2 \kappa({\bi q})
       W_P({\bi q})\ e^{i{\bi q}\cdot {\bi \theta}}\right].
\label{filter2}
\end{eqnarray}
We introduced the window function $W_P(\bi{q})$ to describe the effect of
the detector beam on the convergence. 
It has the form
\begin{eqnarray}
W_P(q)={\int d^2{\bi l} F(l)F(|\bi l+ \bi q|)l^2 C^{\tl P \tl P}_{l} \over
 \int d^2{\bi l}\ F^2(l) l^2 C^{\tl P \tl P}_l}.
\label{windowp}
\end{eqnarray}
We see that $W_P(q) \rightarrow 1$ for $q \rightarrow 0$ independently of
the explicit form of the filter.  Thus for large scales we
reconstruct the convergence power spectrum without beam degradation.
More generally, 
the relation between the observables and underlying convergence is
\begin{equation}
\langle {\cal S}_P({\bi l}) \rangle_{CMB} =
\langle {\cal E}_P({\bi l}) \rangle_{CMB}= 2\kappa({\bi l})W_P(l)
\end{equation}
\begin{figure}[p]
\begin{center}
\vspace*{10cm}
\includegraphics{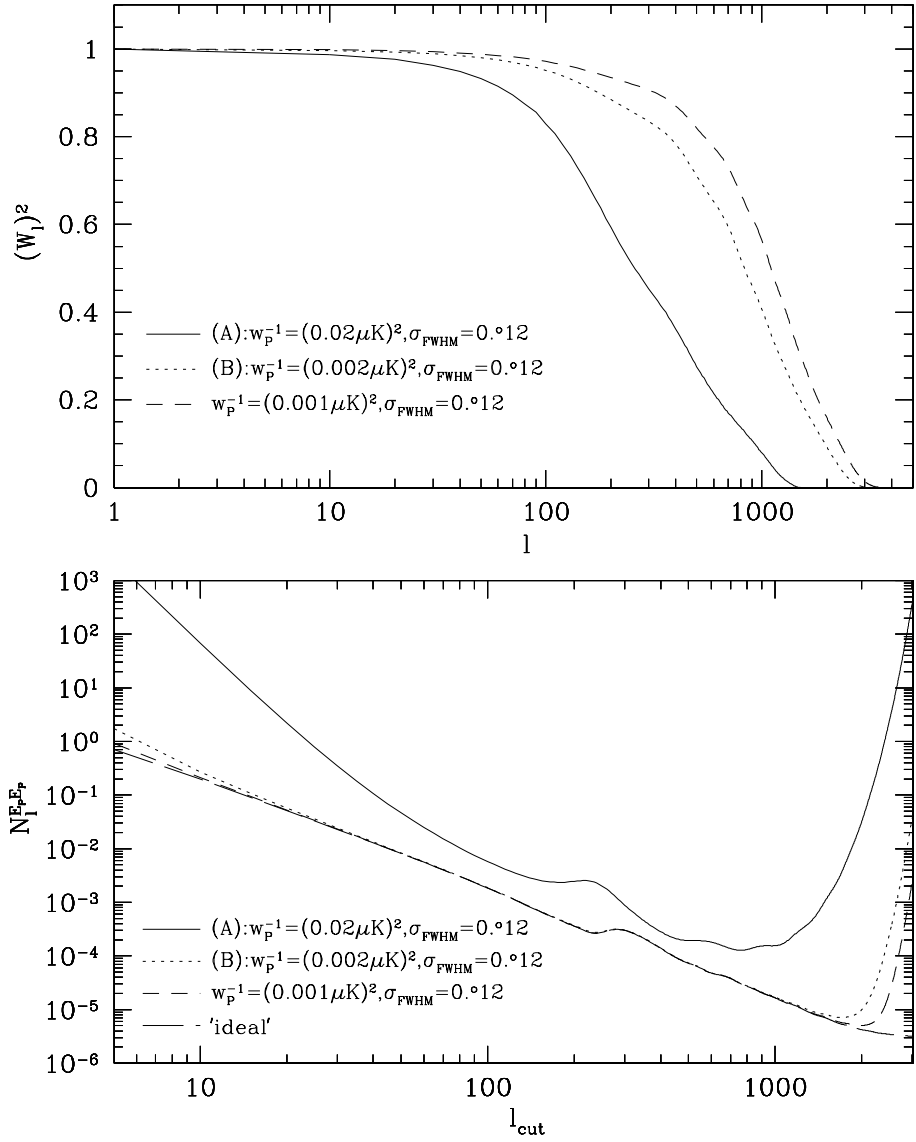}
\end{center}
\caption{In the upper panel square of the window function for  
a few detector configurations using the optimal $l_{cut}$ is shown. 
In the lower panel $N_l^{{\cal E}_P {\cal E}_P}$ is shown as 
a function of $l_{cut}$ using the white 
noise approximation.
Detector (A) is for current Planck proposal.
Increasing the 
detector sensitivity by another order of magnitude, as could be achieved
by post-Planck generation of CMB satellites, would bring the noise 
level close to the ideal case.
}
\label{pwin}
\end{figure}
 
Window for  Planck in the case of polarization and temperature is shown in 
fig.\ \ref{pwin}a. Just like in the case of temperature \cite{lens} we remove 
small scale modes with $l>l_{cut}$ where detector noise exceeds primary 
signal (see below).
At $l \sim 300$ the beam already reduces the sensitivity by 50\%.
Ten times more sensitive detectors
would significantly extend the range of sensitivty and would 
be able to measure polarization spectrum 
up to approximately $l \approx 2000$. The window for such a detector   
is also shown in fig. \ref{pwin}a. 

In the presence of the beam smoothing the
equation (\ref{estim}) 
is rewritten for polarization in the form
\begin{equation}
\langle \hat{C}_l^{{\cal WW}'} \rangle_{CMB} =
4 W^2_P(l) C_l^{\kappa \kappa}+N_l^{\cal WW'},
\label{estim2}
\end{equation}
while for cross-correlation its form is
\begin{equation}
\langle \hat{C}_l^{{\cal WW}'} \rangle_{CMB} =
4 W_T(l)W_P(l) C_l^{\kappa \kappa}+N_l^{\cal WW'}.
\label{estim3}
\end{equation}
Note that to asses the noise power spectrum we need to measure the 
CMB polarization, temperature and cross-correlation spectrum first.

Detector noise affects the results through the increase of the 
CMB correlation length. 
The detector noise spectrum $w^{-1}_P$ is taken as a white noise in the
considered range and is related
to the pixel noise and its solid angle size through
$w^{-1}_P = \sigma^2_{\rm pixel} \Omega_{\rm pixel}$ and similarly for
temperature.
Planck for example is expected 
to have the resolution of $\sigma_{\rm FWHM}=0.12^{\circ}$
and the noise $w^{-1}_P=(0.02)^2 \ (\mu K)^2$.
Individual structures on 
small scales are dominated by noise and 
cannot be observed, hence cannot be used to reconstruct 
convergence so
it is best to remove them. We remove them by filtering the 
data on scale corresponding  
to the scale where noise and signal power spectra are equal, which 
in Fourier space defines some value for $l_{cut}$. 
In fact, for the 
white noise approximation 
the spectra have the form presented in equation (\ref{limitp})
which for polarization gives
\begin{eqnarray}
N_l^{{\cal E}_P {\cal E}_P} =
2\pi {\int_0^{l_{cut}} l^5dl (C_l^{\tl P \tl P}+w^{-1}_P e^{l^2\sigma ^2_b})^2
\over (\int l^3 dlC^{\tl P \tl P}_l)^2}.
\label{limittrue}
\end{eqnarray}
This can be used to determine $l_{cut}$ which minimizes the noise, at 
least on large scales where the approximation is valid. 
In fig.\ \ref{pwin}b are shown a few examples of the
detector configurations as a function of $l_{cut}$. 
For non-zero detector noise the curves have 
minima in $l_{cut}$ above which noise increases
rapidly as expected. Small $l_{cut}$ filters out real CMB 
structure and increase the correlation length, thus increasing 
the level of noise. Once $l_{cut}$ becomes too large the filtering 
scale is too small and small scale power is dominated by noise. This 
does not have any information on the lensing signal and increase the 
overall noise level again. The optimal value for $l_{cut}$ is found 
in between these two regimes and 
agrees well with the value defined where the noise
and signal power spectra in CMB are equal. 

\begin{figure}[p]
\begin{center}
\vspace*{15cm}
\includegraphics{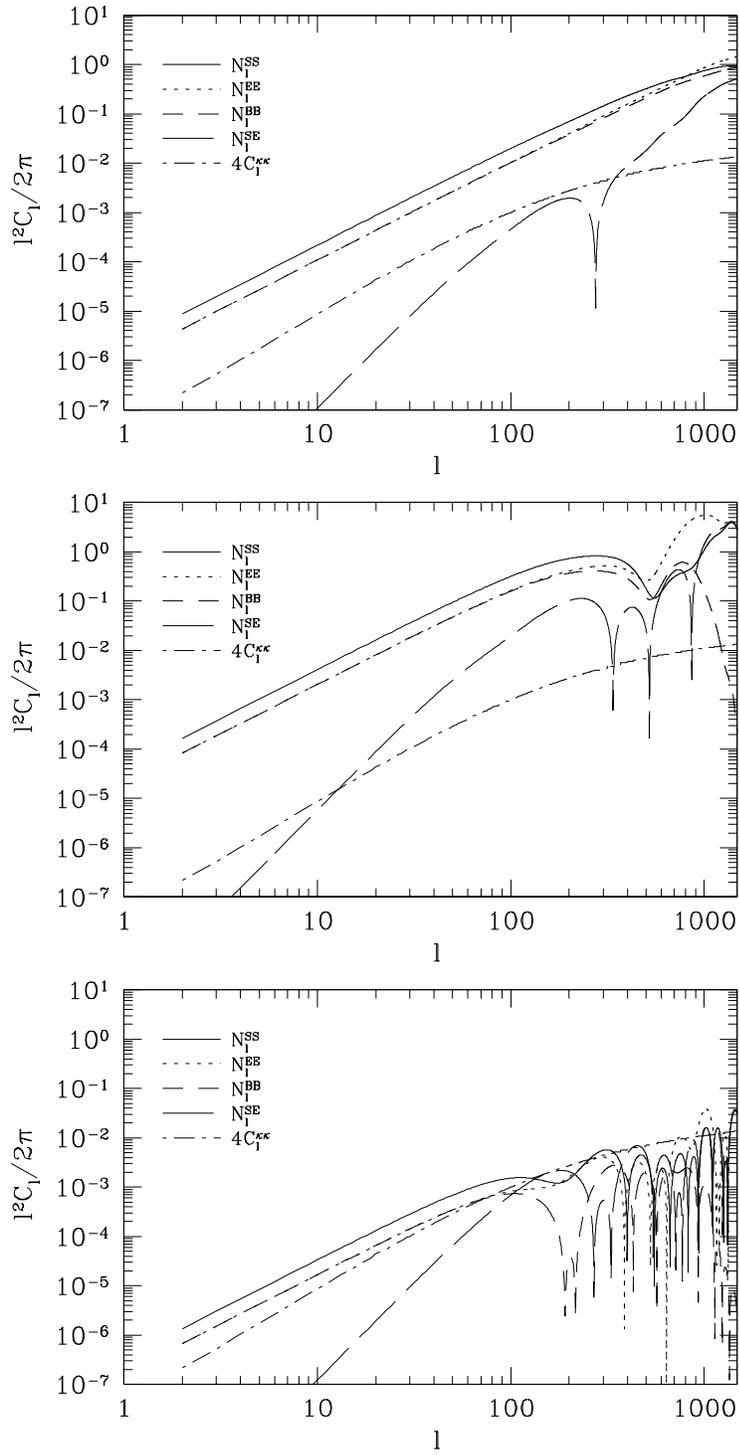}
\end{center}
\caption{Noise power spectra for temperature (upper panel), 
polarization (middle panel) and cross-correlation (lower panel) measurements.
All contain additional noise introduced by Planck detector. }
\label{power}
\end{figure}

\begin{figure}[p]
\begin{center}
\vspace*{12cm}
\includegraphics{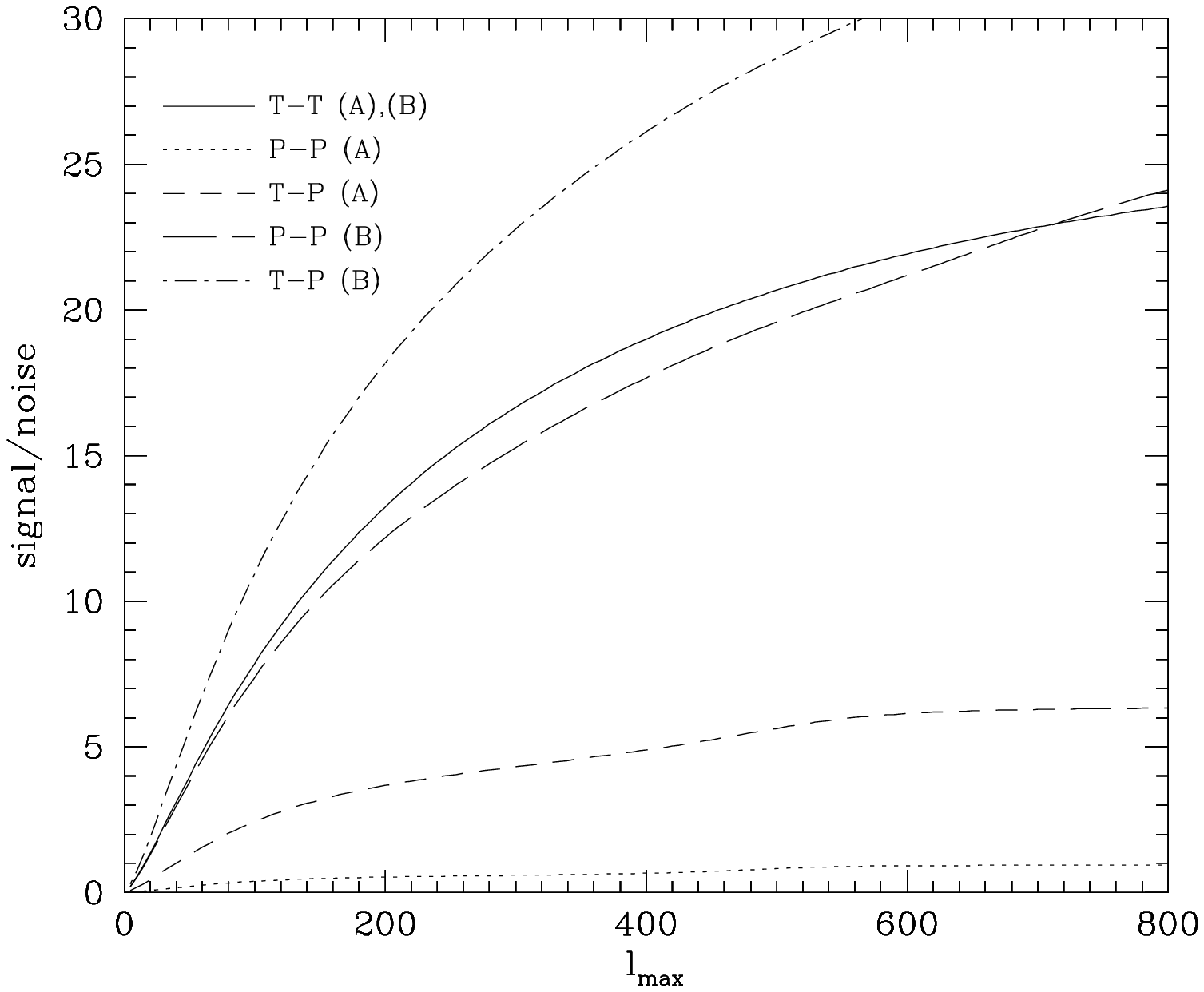}
\end{center}
\caption{ Signal to noise ratio as a function of maximum multipole 
taken into account 
for temperature autocorrelation (T-T), polarization autocorrelation (P-P) and
cross-correlation (T-P). The detector configurations are the same as in 
fig.\ref{pwin}}. For T-T we use $w_T^{-1}=0.002 \mu K$ and 
$\sigma_{FWHM}=0.12^{\circ}$.     
\label{signal}
\end{figure}

Figure\ \ref{power} gives the
expected temperature, polarization and cross-correlation noise spectra 
for Planck
as well as the
expected convergence power spectrum. Noise is always
several times larger than the expected signal so
extraction of $C^{\kappa \kappa}_l$ is possible only in the statistical sense,
by averaging over many mulipoles. In case of 
polarization the noise is about two orders of magnitude larger 
than the signal. This means that for Planck one would need to average 
over $2\times 10^4$ modes to obtain $S/N \sim 1$. This is further 
reduced by the effect of the window. A more accurate estimate of 
expected signal to noise can be obtained by 
combining the information from all multipoles 
and form the optimal estimator
\cite{letlens}. This gives
\begin{equation}
\left({S \over N}\right)_{AB}=
\left( f_{sky}
\sum_l {\ W_A^2(l) \ W_B^2(l) \ \langle\hat{C}^{\kappa \kappa}_l\rangle^2 \over
{\rm Var}\left(\hat{C}^{\kappa \kappa}_l\right)} \right)^{1/2}
\end{equation}
where $f_{sky}$ is the fraction of the 
sky covered, here taken to be 0.7. 
Results are shown in the fig. \ref{signal}.
For  Planck the signal to noise ratio for polarization 
reaches only $(S/N)_{PP} \sim 1$ when we take into account 
about 400 multipoles and does not increase beyond that. 
So the detection of the
convergence by Planck from the polarization measurements 
is not really possible, unless the signal in convergence is 
significantly higher than assumed here. 
The case of the cross-correlation is more promising because of lower 
noise in temperature. Here we find it is possible to
get $(S/N)_{TP} \sim 5$ from $l<400$. 
More promising numbers are found if we 
increase
the detector sensitivity by a factor of several, which could only be 
achieved from 
the next generation of CMB satellites dedicated to polarization. 
Improving the sensitivity by a factor of 10 
makes $S/N$ from polarization 
comparable to the one from temperature, while the cross-correlation 
$S/N$ exceeds both and would provide most of the information.

\section{Conclusions}

We have analysed the prospects of measuring the projected dark matter 
power spectrum up to $z \sim 1100$ from the distortions induced in 
the CMB polarization. Any such measurement would be extremely 
difficult both from the temperature and polarization because of the statistical 
nature of such a detection. On the other hand, it would provide a 
direct detection of dark matter clustering on very large scales and 
high redshifts, possibly not attainable by any other method. For 
this reason it is important that any such detection
from temperature fluctuations,
described in \cite{lens}, is independently confirmed. 
Such a confirmation is possible with polarization, with which one 
would be provided 
in total with 10 power spectra of convergence, all of which 
have to agree with each other. 

Because polarization is an order of magnitude smaller than 
temperature fluctuations its detection will be rather limited 
with the planned CMB experiments including Planck. 
The expected
signal to noise for convergence power spectrum for 
Planck is around unity using 
polarization information only, compared to 15-25 from 
temperature. Cross-correlation could be more promising and 
could provide $S/N \sim 5$ or a factor of two larger or smaller
depending on the actual amount of power in convergence. The 
real promise of this method lies in the contemplated 
post-Planck experiments dedicated to polarization.
Such an experiment
would need to cover
a significant fraction of the sky to measure the 
projected power spectrum on large scales.
If the sensitivity of detectors is reduced by another 
order of magnitude over Planck then measurement of dark matter 
power spectrum with polarization becomes very significant and 
would extend the 
amount of information from temperature 
alone. It would also allow for many cross-checks of the results to 
reduce the possible systematics. 

\smallskip
J.G. was supported by grant 2-P03D-022-10 from Polish State Committee for 
Scientific Research. 
U.S. ackowledges the support of NASA grant NAG5-8084.
M.Z. is supported by NASA through Hubble Fellowship grant
HF-01116.01-98A from STScI,
operated by AURA, Inc. under NASA contract NAS5-26555.
J.G. and U.S. 
would like to thank Max-Planck Institut f\" ur Astrophysik, Garching,
for hospitality during the visits.

\appendix
\section{Correlation functions formulas}

Here we present correlation functions of derivatives of the CMB temperature 
and polarization fields. Due to the rotational invariance 
of the variables ${\cal S}_T$, ${\cal S}_P$,
${\cal E}_T$, ${\cal E}_P$ it is enough to consider correlations between two 
directions separated by $\theta$ along the $x$ direction. During derivation of 
the correlation functions we make use of the integral representation of  
Bessel functions and the CMB spectra defined in \cite{urospol}. 
Weighted spectra $C_i(\theta)$ are defined as 
\begin{eqnarray}
C_i^{\tl A \tl B} =
  \int {l^3dl \over 2\pi} C_l^{\tl A \tl B} \ J_i(l \theta), 
\label{weighted}
\end{eqnarray}
where $\tl A$ and $\tl B$ stand for $\tl T$ or $\tl P$.  
Some calculations yields the following correlation functions 
\begin{equation} \; \label{appform} \end{equation}
\begin{eqnarray*}
C_{xx}^{QQ}(\theta)&\equiv&\langle \tl Q_x(0) \tl Q_x(\theta) \rangle_{CMB}
\\
&=&(2\pi)^{-2}\int d^2{\bi l}\ 
e^{i l\cdot \theta \cos\phi_l}\ l^2 \cos^2\phi_l C^{\tl P \tl P}_l \cos^2 2\phi_l \nonumber \\
&=&\int {l dl \over 2 \pi}\ 
l^2C^{\tl P \tl P}_l {1 \over 4} \left[ J_0(l\theta) - 3/2 J_2(l\theta) + J_4(l\theta) - 1/2  J_6(l\theta) \right] \\
&\equiv&{1 \over 4}[C_0^{\tl P \tl P}(\theta) - 3/2 C_2^{\tl P \tl P}(\theta) + C_4^{\tl P \tl P}(\theta) - 1/2  C_6^{\tl P \tl P}(\theta)] \nonumber \\
C_{yy}^{QQ}(\theta)&\equiv&\langle \tl Q_y(0) \tl Q_y(\theta) \rangle_{CMB} \nonumber \\
&=&(2\pi)^{-2}\int d^2{\bi l}\ 
e^{i l\cdot \theta \cos\phi_l}\ l^2 \sin^2\phi_l C^{\tl P \tl P}_l \cos^2 2\phi_l \nonumber \\
&=&\int {l dl \over 2 \pi}\ 
l^2C^{\tl P \tl P}_l {1 \over 4} \left[ J_0(l\theta) + 3/2 J_2(l\theta) + J_4(l\theta) + 1/2 J_6(l\theta) \right] \\ 
&=&{1 \over 4}[C_0^{\tl P \tl P}(\theta)+ 3/2 C_2^{\tl P \tl P}(\theta) + C_4^{\tl P \tl P}(\theta) + 1/2 C_6^{\tl P \tl P}(\theta)] \nonumber \\
C_{xy}^{QQ}(\theta)&\equiv&\langle \tl Q_x(0) \tl Q_y(\theta) \rangle_{CMB} \nonumber \\
&=&(2\pi)^{-2}\int d^2{\bi l}\ 
e^{i l\cdot \theta \cos\phi_l}\ l^2 \sin\phi_l \cos\phi_l C^{\tl P \tl P}_l \cos^2 2\phi_l  \\
&=& 0  \nonumber \\
\\
C_{xx}^{UU}(\theta)&\equiv&\langle \tl U_x(0) \tl U_x(\theta) \rangle_{CMB} \nonumber \\
&=&(2\pi)^{-2}\int d^2{\bi l}\ 
e^{i l\cdot \theta \cos\phi_l}\ l^2 \cos^2\phi_l C^{\tl P \tl P}_l \sin^2 2\phi_l \nonumber \\
&=&\int {l dl \over 2 \pi}\ 
l^2C^{\tl P \tl P}_l {1 \over 4} \left[ J_0(l\theta) - 1/2 J_2(l\theta) - J_4(l\theta) + 1/2 J_6(l\theta) \right] \\
&=&{1 \over 4}[C_0^{\tl P \tl P}(\theta) - 1/2 C_2^{\tl P \tl P}(\theta) - C_4^{\tl P \tl P}(\theta) + 1/2 C_6^{\tl P \tl P}(\theta)] \nonumber \\
C_{yy}^{UU}(\theta)&\equiv&\langle \tl U_y(0) \tl U_y(\theta) \rangle_{CMB} \nonumber \\
&=&(2\pi)^{-2}\int d^2{\bi l}\ 
e^{i l\cdot \theta \cos\phi_l}\ l^2 \sin^2\phi_l C^{\tl P \tl P}_l \sin^2 2\phi_l \nonumber \\
&=&\int {l dl \over 2 \pi}\ 
l^2C^{\tl P \tl P}_l {1 \over 4} \left[ J_0(l\theta) + 1/2 J_2(l\theta) - J_4(l\theta) - 1/2 J_6(l\theta) \right]  \\ 
&=&{1 \over 4}[C_0^{\tl P \tl P}(\theta)+ 1/2 C_2^{\tl P \tl P}(\theta) - C_4^{\tl P \tl P}(\theta) - 1/2 C_6^{\tl P \tl P}(\theta)] \nonumber \\
C_{xy}^{UU}(\theta)&\equiv&\langle \tl U_x(0) \tl U_y(\theta) \rangle_{CMB} \nonumber \\
&=&(2\pi)^{-2}\int d^2{\bi l}\ 
e^{i l\cdot \theta \cos\phi_l}\ l^2 \sin\phi_l \cos\phi_l C^{\tl P \tl P}_l \sin^2 2\phi_l \\
&=& 0 \nonumber \\
\\
C_{xx}^{TQ}(\theta)&\equiv&
\langle \tl T_x(0) \tl Q_x(\theta) \rangle_{CMB} \nonumber \\
&=&(2\pi)^{-2}\int d^2{\bi l}\ 
e^{i l\cdot \theta \cos\phi_l}\ l^2 \cos^2\phi_l C^{\tl T \tl P}_l \cos 2\phi_l \nonumber \\
&=&\int {l dl \over 2 \pi}\
l^2\ C^{\tl T \tl P}_l {1 \over 4} \left[ J_0(l\theta) - 
2 J_2(l\theta) + J_4(l\theta) \right] \\
&=&{1 \over 4}[ C_0^{\tl T \tl P}(\theta) -2 C_2^{\tl T \tl P}(\theta)+ C_4^{\tl T \tl P}(\theta)] \nonumber \\
C_{yy}^{TQ}(\theta)&\equiv&
\langle \tl T_y(0) \tl Q_y(\theta) \rangle_{CMB} \nonumber \\
&=&(2\pi)^{-2}\int d^2{\bi l}\ e^{i l\cdot \theta \cos\phi_l}\ l^2
\sin^2\phi_l C^{\tl T \tl P}_l \cos 2\phi_l \nonumber \\
&=&-\int {l dl \over 2 \pi}\ l^2\ C^{\tl T \tl P}_l
{1 \over 4} \left[ J_0(l\theta) + 2 J_2(l\theta) + J_4(l\theta) \right] \\
&=&-{1 \over 4}[ C_0^{\tl T \tl P}(\theta)+2 C_2^{\tl T \tl P}(\theta)+C_4^{\tl T \tl P}(\theta)] \nonumber \\
C_{xy}^{TQ}(\theta)&\equiv&
\langle \tl T_x(0) \tl Q_y\theta) \rangle_{CMB} \nonumber \\
&=&(2\pi)^{-2}\int d^2{\bi l}\ e^{i l\cdot \theta \cos\phi_l}\ l^2
\sin\phi_l \cos\phi_l C^{\tl T \tl P}_l \cos 2\phi_l  \\
&=& 0 \nonumber \\
\\
C_{xx}^{TU}(\theta)&\equiv&
\langle \tl T_x(0) \tl U_x(\theta) \rangle_{CMB} \nonumber \\
&=&(2\pi)^{-2}\int d^2{\bi l}\ e^{i l\cdot \theta \cos\phi_l}\ l^2
\cos^2\phi_l C^{\tl T \tl P}_l \sin 2\phi_l  \\
&=& 0 \nonumber \\
C_{yy}^{TU}(\theta)&\equiv&
\langle \tl T_y(0) \tl U_y(\theta) \rangle_{CMB} \nonumber \\
&=&(2\pi)^{-2}\int d^2{\bi l}\ e^{i l\cdot \theta \cos\phi_l}\ l^2
\sin^2\phi_l C^{\tl T \tl P}_l \sin 2\phi_l  \\
&=& 0 \nonumber \\
C_{xy}^{TU}(\theta)&\equiv&
\langle \tl T_x(0) \tl U_y(\theta) \rangle_{CMB} \nonumber \\
&=&(2\pi)^{-2}\int d^2{\bi l}\ e^{i l\cdot \theta \cos\phi_l}\ l^2
\sin\phi_l \cos\phi_l C^{\tl T \tl P}_l \sin 2\phi_l \nonumber \\
&=&\int {l dl \over 2 \pi}\ l^2\ C^{\tl T \tl P}_l
{1 \over 4} \left[ J_0(l\theta) - J_4(l\theta) \right] \\
&=&{1 \over 4}[ C_0^{\tl T \tl P}(\theta) - C_4^{\tl T \tl P}(\theta)] \nonumber \\
\\
C_{xx}^{QU}(\theta)&\equiv&
\langle \tl Q_x(0) \tl U_x(\theta) \rangle_{CMB} \nonumber \\
&=&(2\pi)^{-2}\int d^2{\bi l}\ e^{i l\cdot \theta \cos\phi_l}\ l^2
\cos^2 \phi_l C^{\tl P \tl P}_l \sin 2\phi_l \cos 2\phi_l  \\
&=& 0 \nonumber \\
C_{yy}^{QU}(\theta)&\equiv&
\langle \tl Q_y(0) \tl U_y(\theta) \rangle_{CMB} \nonumber \\
&=&(2\pi)^{-2}\int d^2{\bi l}\ e^{i l\cdot \theta \cos\phi_l}\ l^2
\sin^2 \phi_l C^{\tl P \tl P}_l \sin 2\phi_l \cos 2\phi_l  \\
&=& 0 \nonumber \\
C_{xy}^{QU}(\theta)&\equiv&
\langle \tl Q_x(0) \tl U_y(\theta) \rangle_{CMB} \nonumber \\
&=&(2\pi)^{-2}\int d^2{\bi l}\ e^{i l\cdot \theta \cos\phi_l}\ l^2
\sin\phi_l \cos\phi_l C^{\tl P \tl P}_l \sin 2\phi_l \cos 2\phi_l \nonumber \\
&=& - \int {l dl \over 2 \pi}\ l^2\ C^{\tl P \tl P}_l
{1 \over 8} \left[ J_2(l\theta) - J_6(l\theta) \right] \\
&=&-{1 \over 8}[ C_2^{\tl P \tl P}(\theta) - C_6^{\tl P \tl P}(\theta)]. \nonumber
\end{eqnarray*}

\end{document}